\documentclass[conference]{IEEEtran}

\ifCLASSINFOpdf

\else

\fi

\usepackage{cite}
\usepackage{amsmath,amssymb,amsfonts}
\usepackage{algorithmic}
\usepackage{graphicx}
\usepackage{textcomp}
\usepackage{xcolor}
\begin{document}

\title{FinGAN: An Interpretable RSS Generation Network for Scalable Fingerprint Localization
}

\author{Jiaming~Zhang\IEEEauthorrefmark{2}, Jiajun~He\IEEEauthorrefmark{2}, Jie~Zhang\IEEEauthorrefmark{2}, Okan~Yurduseven\IEEEauthorrefmark{2}
\\
\IEEEauthorrefmark{2}Centre for Wireless Innovation, Queen's University Belfast, Belfast, Belfast BT3 9DT, UK

\thanks{This work was supported by the Leverhulme Trust under Research Leadership Award RL-2019-019.}}

\markboth{Journal of \LaTeX\ Class Files,~Vol.~14, No.~8, August~2015}%
{Shell \MakeLowercase{\textit{et al.}}: Bare Demo of IEEEtran.cls for IEEE Journals}

\maketitle

\begin{abstract}
This work introduces FinGAN, a robust received signal strength (RSS) data generator designed to expand RSS fingerprint datasets. Compared to existing generative adversarial models that either rely on known reference positions (RPs) or depend on predefined priors, FinGAN learns the latent information between RPs and RSS values by maximizing the mutual information between the generated RSS data and the RPs, enabling an end-to-end RSS generation directly from RPs. This allows us to accurately generate RSS data for previously unmeasured RPs. Both quantitative and qualitative evaluations demonstrate that FinGAN produces synthetic RSS data closely aligned with real RSS sample collected from the on-site experiment, preserving localization performance comparable to that achieved with complete real-world datasets. To further validate its generalizability, FinGAN is also trained and evaluated on open-source datasets from three typical office environments,and the results demonstrate consistent performance across different scenarios.
\end{abstract}

\begin{IEEEkeywords}
Fingerprint, generative adversarial learning, localization, received signal strength 
\end{IEEEkeywords}

\section{Introduction}
\label{sec:introduction}
\IEEEPARstart{P}{ositioning} using RSS has gained considerable interest due to its simplicity, low cost, and ease of integration into existing wireless systems, supporting applications such as navigation, tracking, and surveillance \cite{so-loc}. However, its accuracy is typically lower than time- or angle-based methods because RSS is highly sensitive to channel fluctuations. Motivated by this limitations, fingerprinting-based localization has been developed, offering accurate target estimation even under severe non-line-of-sight (NLoS) conditions. In general, fingerprinting operates in two phases: (i) offline, where a database of RSS measurements at known locations is constructed, and (ii) online, where new RSS samples are matched against the database to estimate target locations. Despite its high accuracy, several challenges remain \cite{RSSChara}:

\begin{enumerate}
    \item \textit{RSS variation}: Values fluctuate due to radio unit (RU) mobility, transmit power changes, environmental modifications, and long-term signal degradation \cite{heTMC2017}, which can significantly render databases outdated.  
    
    \item \textit{Grid size}: Finer training grids improve accuracy but greatly increase database construction time and cost \cite{he2022RSS}.  
    
    \item \textit{Data scarcity}: Limited RSS observations degrade performance \cite{liu2021tensorGAN}, and large-scale databases are especially costly in wide-area or outdoor scenarios.  
\end{enumerate}

These issues motivate the development of an RSS generator to reduce database construction overhead while ensuring reliable localization.

To reduce reliance on large-scale RSS collection, Gaussian Process Regression (GPR) has been used to model the statistical distribution of RSS based on known RPs \cite{8464281}, enabling the generation of additional samples for localization. However, GPR-based methods suffer from limited scalability and ability to capture complex RSS patterns \cite{zou2020robot}. To mitigate temporal variability that can render databases outdated, He \textit{et al.} \cite{heTMC2017} proposed a GPR-based updating scheme, while an RSS calibration mechanism was later introduced to address device heterogeneity and user mobility \cite{heTMCcal}. Recently, generative adversarial networks (GANs) have emerged as a promising alternative for synthesizing RSS fingerprints. Serreli \textit{et al.} \cite{serreli2024generative} combined a GAN with a Bayesian neural network for classification, while Boulis \textit{et al.} \cite{boulis2021data} employed conditional GANs (cGANs) to condition RSS generation on measured RPs. To capture temporal dynamics, Junoh \textit{et al.} \cite{10443392} incorporated long short-term memory (LSTM) modules into a cGAN. However, these methods are constrained to RPs already present in the database and cannot generalize to \textit{unmeasured} RPs. To address this limitation, Chen \textit{et al.} \cite{9812625} proposed a GAN framework that jointly generates RPs and RSS from random latent vectors. While this enables synthesis at unseen RPs, it lacks explicit control over generation conditioned on specific RPs, limiting the precision of the RP–RSS mapping. In addition, GANs have been applied to refine RSS estimation; for instance, Zou \textit{et al.} \cite{zou2020robot} introduced GPR-GAN, a cascaded model where a GAN enhances GPR-generated RSS. Although this allows conditioning on both measured and unmeasured RPs, the approach remains restricted by the limited capacity and kernel sensitivity of GPR.

\begin{figure}[t]
    \centering
    \includegraphics[width=0.7\linewidth]{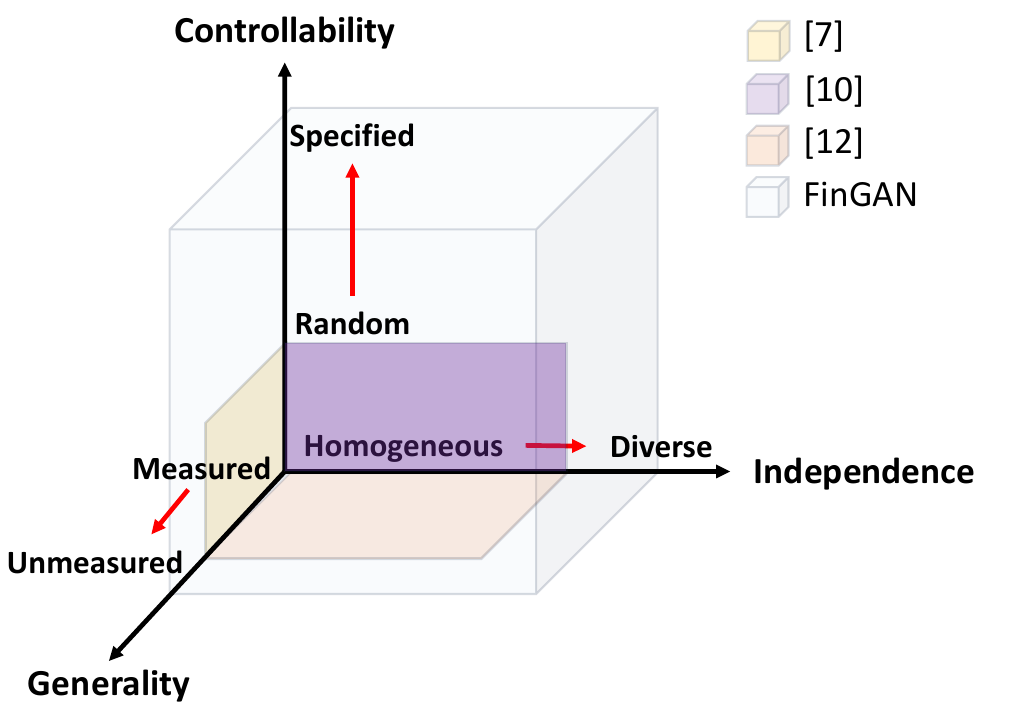}
    \caption{Contribution of this work.}
    \label{fig:figcon}
    \vspace{-15pt}
\end{figure}

Inspired by Information Maximizing GAN (InfoGAN) \cite{chen2016infogan}, an RSS generator, termed FinGAN, is proposed. While InfoGAN has been widely studied in computer vision, its direct application to RSS fingerprint localization is non-trivial due to the non-stationary, highly noisy nature of wireless signals. FinGAN integrates generative modelling with interpretable latent representations to reduce fingerprinting effort and improve localization accuracy. This design provides both theoretical novelty in RP–RSS mutual information modelling and practical novelty in reducing costly fingerprint data collection. Different from the standard InfoGAN, which learns disentangled latent factors in vision tasks, FinGAN introduces a tailored variant for wireless localization: (i) embedding RP coordinates as interpretable latent codes, (ii) employing an auxiliary spatial network (AuxNet) to capture RP–RSS dependencies, and (iii) enabling controllable synthesis at unseen RPs. These adaptations make FinGAN fundamentally different from vanilla InfoGAN and specifically suited to noisy, non-stationary RSS data. As shown in Fig. \ref{fig:figcon}, the contributions of this paper are listed as follows:

\begin{enumerate}
    \item \textit{Controllability}: FinGAN synthesizes RSS data at arbitrary RPs by conditioning generation on RP indices. An auxiliary network (AuxNet) embedded in the discriminator captures spatial features and RSS patterns, enabling realistic data generation beyond random or limited RPs.
    
    \item \textit{Independence}: FinGAN directly learns the RP--RSS mapping in an end-to-end manner. By maximizing mutual information between inputs and outputs, it removes dependency on prior models (e.g., GPR) and enhances robustness and generalization.
    
    \item \textit{Generality}: FinGAN models the joint distribution of RSS and RPs, supporting RSS generation at unmeasured locations. Adversarial training yields high-fidelity synthetic data that closely follow real measurements.
\end{enumerate}

\section{Problem Formulation}
\label{sec:Problem Formulation}
\subsection{RSS Measurement}
In a two-dimensional (2-D) localization setting, the objective is to estimate the target position $\mathbf{p} = [x \ y]^T$ using $L$ RUs. Denoting the transmit power of the target (a mobile device) be $P_{t}$, the received signal power $P_{r,l}$ between the target and $l$-th RU ($\mathbf{p}_l$) is computed by \cite{he2022RSS}: 
\begin{equation}
    P_{r,l} = P_t K_l h_{l}d_{l}^{-\alpha},
\label{eq:power}
\end{equation}
\begin{equation}
    d_l = \| \mathbf{p} - \mathbf{p}_l \|.
\label{eq:power_c'}
\end{equation}
where $\|\cdot\|$ denotes the Euclidean norm, $K_l$ is the antenna gain, $h_{l}$ denotes the channel fading gain, while $\alpha$ is the path-loss exponent. By applying a natural logarithm to both sides of \eqref{eq:power}, the RSS observed at the $l$-th RU can be written as:
\begin{equation}
    r_{\text{RSS},l} = \ln P_t - \ln h_l - \ln K_l = -\alpha\ln(d_l) + n,
    \label{eq:rss}
\end{equation}
where $n$ is the additive noise. while fading and noise introduce stochastic fluctuations. In practice, however, RSS measurements are highly unstable due to multipath, non-line-of-sight (NLoS) propagation, and blockage effects, which substantially degrade the accuracy of multilateration-based localization \cite{5739089, XIONG2024104653}. To overcome these challenges, fingerprinting-based localization leverages the statistical distribution of RSS values, rather than a strict geometric model, making it a robust alternative in mixed line-of-sight (LoS) and NLoS environments.

\subsection{Fingerprinting Localization}

\begin{figure}[t!]
\centerline{\includegraphics[width=0.8\columnwidth]{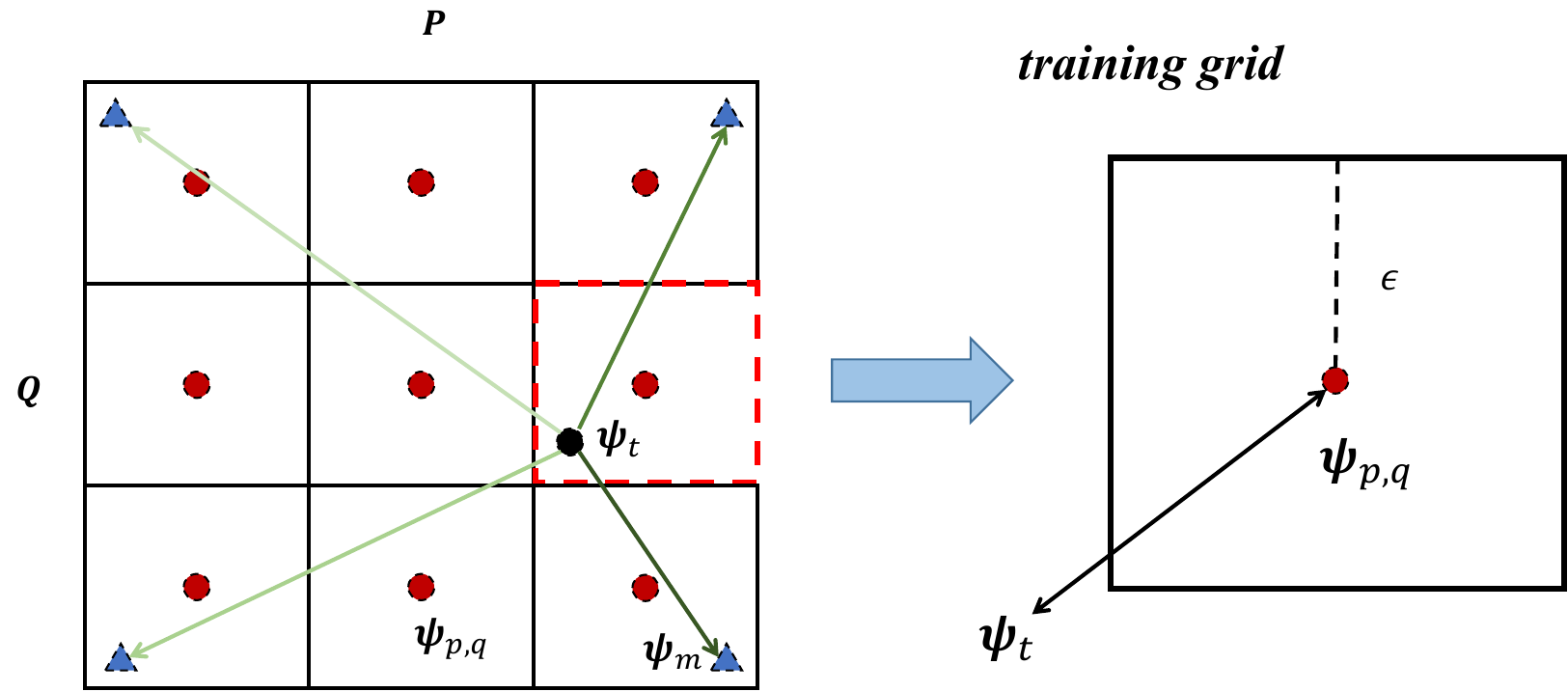}}
\caption{Illustration of training grids.}
\label{fig-system model}
\vspace{-15pt}
\end{figure}

To address the problem encountered in the multilateration, fingerprinting can be applied. The fingerprinting algorithm includes two major phases: 1) offline and 2) online phases. Fig. \ref{fig-system model} illustrates that the localization area is divided into $L\times W$ training grids with resolution $\epsilon$. The center point of training grid is denoted by: $\mathbf{p}_{l,w} = [x_{l,w}, y_{l,w}]^T$, where $l = 1, 2, \ldots, L$ and $w = 1, 2, \ldots, W$. Assuming that $N$ fingerprints are collected by the $m$-th RU, the vector form of the fingerprints is given by \cite{he-LTE}:
\begin{equation}
\begin{split}
    \mathbf{f}_{m} = [r_{{\rm{RSS}},m}^{1},\ldots,r_{{\rm{RSS}},m}^N]^T, \\
\end{split}
\end{equation}
where $\mathbf{f}_{m}$ contains both the LoS and NLoS measurements. The recorded RSSs are then utilized to construct the fingerprint database, which will be leveraged to match the online measurements for location estimation. Thus, fingerprinting localization is generally regarded as a pattern-matching problem. For example, a similarity kernel function, such as the cosine similarity \cite{cosinesim}, can be utilized to find the most similar fingerprints in the database. However, directly matching online fingerprints with the database significantly decreases the localization performance. Thus, data pre-processing and filtering techniques, such as the Kalman and particle filters, should be applied to remove abnormal and unreliable RSS values. Furthermore, the performance of the fingerprinting algorithm depends on the resolution of the training grid, while increasing $\epsilon$ significantly lowers the computational efficiency of the algorithm. This trade-off highlights the need for alternative approaches that directly exploit the statistical and geometric properties of RSS, rather than relying solely on dense fingerprint databases.

\begin{figure*}[t]
    \centering
    \includegraphics[width=0.8\textwidth]{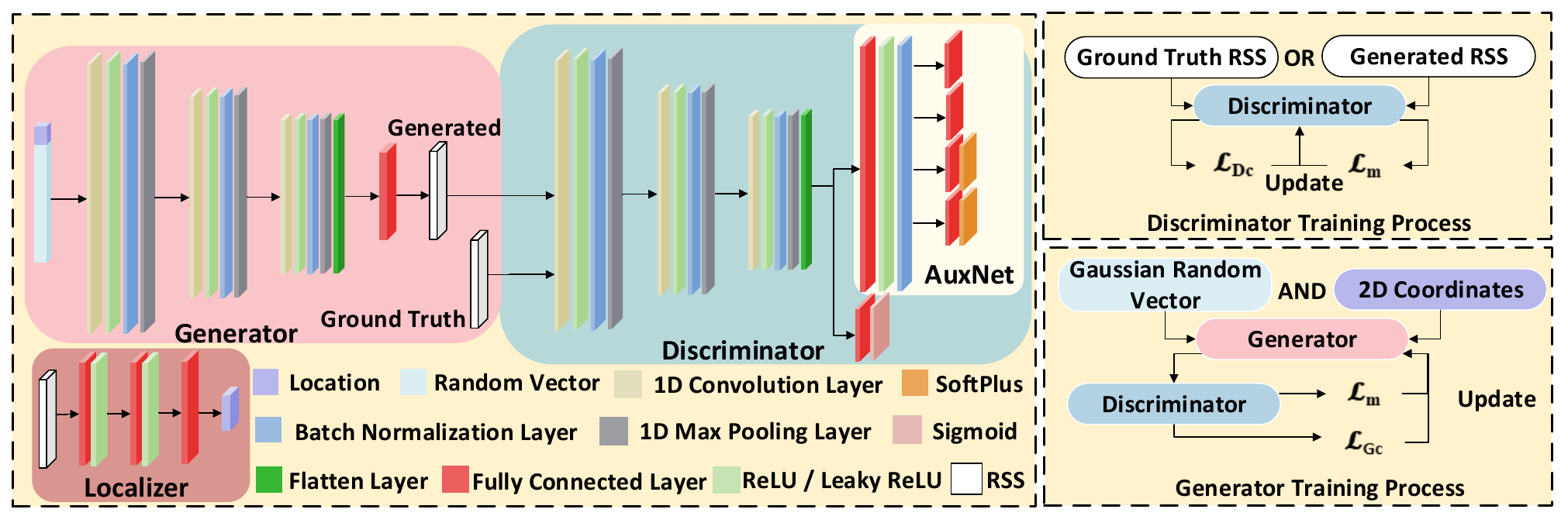}
    \caption{The architecture and training process of FinGAN, including the localizer network.}
    \label{fig:FinGAN_archi}
    \vspace{-15pt}
\end{figure*}

\section{FinGAN-based RSS Generator}
\label{sec:FinGAN-based RSS Generator}
\subsection{Motivation}
As shown in (\ref{eq:power_c'}) and (\ref{eq:rss}), the path-loss model implies that RSS decreases monotonically with the logarithm of the RP–RU distance, indicating that RSS inherently encodes geometric information. This motivates a generative approach where the RP serves as a conditioning variable. A natural choice is the cGAN, which generates data conditioned on labels such as RPs; however, cGAN assumes label independence and requires consistent labels across training and testing, limiting flexibility. To address this, FinGAN replaces discrete labels with interpretable latent codes sampled from continuous distributions. The generator thus receives both random noise and latent codes: noise captures overall variability, while latent codes modulate specific attributes of the outputs. Different from fixed labels, these latent codes allow flexible inference beyond the training set. FinGAN therefore pursues two objectives: (i) learning a mapping from noise to realistic samples, and (ii) maximizing the mutual information between latent codes and generated data.

From an information-theoretic perspective, the mutual information $I(v_1;v_2)$ measures the information content related to a variable $v_1$ which can be learnt from the other variable $v_2$. In the context of this work, the mutual information is defined as: $I(\mathbf{c};\text{G}(\mathbf{z}|\mathbf{c})))$, where $G(\mathbf{z}|\mathbf{c})$ indicates synthetic RSS samples, and $\mathbf{c}$ represents the 2-D RPs used as the latent code. $I(\mathbf{c};\text{G}(\mathbf{z}|\mathbf{c})))$ reflects the dependency between the latent codes and generated samples. The mutual information between $\mathbf{c}$ and $\text{G}(\mathbf{z}|\mathbf{c}))$ can be calculated by \cite{he2022RSS}: $I(\mathbf{c};\text{G}(\mathbf{z}|\mathbf{c}))) = H(\mathbf{c}) - H(\mathbf{c}|\text{G}(\mathbf{z}|\mathbf{c})))$,
where $H(\cdot)$ denotes the entropy. Considering $H(\mathbf{c})$ is a constant, to maximize the mutual information $I(\mathbf{c};\text{G}(\mathbf{z}|\mathbf{c})))$, $H(\mathbf{c}|\text{G}(\mathbf{z}|\mathbf{c})))$ should be minimize. For convenience, use $\mathbf{g}$ to represent $\text{G}(\mathbf{z}|\mathbf{c})$. The mutual information can be further transformed as follows:
\begin{equation} 
    \begin{split}
    I(\mathbf{c};\text{G}(\mathbf{z}|\mathbf{c}))) = {}&I(\mathbf{c};\mathbf{g}))=H(\mathbf{c}) - H(\mathbf{c}|\mathbf{g}))\\
    ={}&H(\mathbf{c}) + \mathbb{E}_{\mathbf{g}\sim P(\mathbf{g}),\mathbf{c}\sim P(\mathbf{c}|\mathbf{g})}[\text{log} P(\mathbf{c}|\mathbf{g})].
    \end{split}
\end{equation}
Since the posterior $P(\mathbf{c}|\mathbf{g})$ is not available to compute directly, an auxiliary distribution $Q(\mathbf{c}|\mathbf{g})$ is proposed to approximate the value of $P(\mathbf{c}|\mathbf{g})$. Employing variational lower bound technique \cite{barber2004algorithm}, it is given by:
\begin{equation} 
    \begin{split}
    I(\mathbf{c};\mathbf{g})) ={}&H(\mathbf{c}) + \mathbb{E}_{\mathbf{g}\sim P(\mathbf{g}),\mathbf{c}\sim P(\mathbf{c}|\mathbf{g})}[\text{log} P(\mathbf{c}|\mathbf{g})]\\
    ={}&H(\mathbf{c}) + \mathbb{E}_{\mathbf{g}\sim P(\mathbf{g})}[D_{KL}(P(\mathbf{c}|\mathbf{g})||Q(\mathbf{c}|\mathbf{g}))]+\\
    {}&\mathbb{E}_{\mathbf{g}\sim P(\mathbf{g}),\mathbf{c}\sim P(\mathbf{c}|\mathbf{g})}[\text{log}Q(\mathbf{c}|\mathbf{g})]\\
    \geq {}&H(\mathbf{c}) + \mathbb{E}_{\mathbf{g}\sim P(\mathbf{g}),\mathbf{c}\sim P(\mathbf{c}|\mathbf{g})}[\text{log}Q(\mathbf{c}|\mathbf{g})],
    \end{split}
\label{eq_mu}
\end{equation}
where $D_{KL}$ is the Kullback-Leibler divergence (KLD), while the lower bound can then be maximized as a proxy for the mutual information. We next introduce the use of the KLD to design the loss function for the proposed problem.

\subsection{Information Maximization Loss Function}
The adversarial objectives for the discriminator $\mathcal{L}_{Dc}$ and generator $\mathcal{L}_{Gc}$ are given by
\begin{equation} 
    \begin{split}
    \mathcal{L}_{Dc} ={}-&\mathbb{E}_{\mathbf{z}\sim P_\mathbf{z}}[(\log (1-\text{D}(\text{G}(\mathbf{z}|\mathbf{c})|\mathbf{c}))]\\
                      & -\mathbb{E}_{\mathbf{x}\sim P_\mathbf{x}}[\log (\text{D}(\mathbf{x}|\mathbf{c}))],\\
    \end{split}
\label{eq_lossD}
\end{equation}
\begin{equation} 
    \begin{split}
    \mathcal{L}_{Gc} =  {}&- \mathbb{E}_{\mathbf{z}\sim P_\mathbf{z}}[\log (\text{D}(\text{G}(\mathbf{z}|\mathbf{c})|\mathbf{c})))],\\
    \end{split}
\label{eq_lossG}
\end{equation}
where $\mathbb{E}$ denotes mathematical expectation, $\mathbf{x}$ indicates the real RSS samples, and $D(\cdot)$ denotes the probability that the input is authentic. Therefore, the total loss functions for FinGAN generator $\mathcal{L}_{G}$ and discriminator $\mathcal{L}_{D}$ are given by:
\begin{equation}
\begin{split}
    \mathcal{L}_{G} = {}&\mathcal{L}_{Gc} - \lambda[\mathbb{E}_{\mathbf{g}\sim P(\mathbf{g}),\mathbf{c}\sim P(\mathbf{c}|\mathbf{g})}[\text{log}Q(\mathbf{c}|\mathbf{g})]]\\
    ={}&\mathcal{L}_{Gc} - \lambda\mathcal{L}_{m},\\
\end{split}
\end{equation}
\begin{equation}
\begin{split}
    \mathcal{L}_{D} = {}&\mathcal{L}_{Dc} - \lambda[\mathbb{E}_{\mathbf{g}\sim P(\mathbf{g}),\mathbf{c}\sim P(\mathbf{c}|\mathbf{g})}[\text{log}Q(\mathbf{c}|\mathbf{g})]]\\
    ={}&\mathcal{L}_{Dc} - \lambda\mathcal{L}_{m},\\
\end{split}
\end{equation}
where $\lambda$ is a hyper-parameter that balances the adversarial and information-maximization objectives. In this work, $\lambda = 1$ is adopted, as it was observed to achieve a robust balance between adversarial and mutual information objectives without requiring further tuning.

\subsection{FinGAN Architecture}
As shown in Fig. \ref{fig:FinGAN_archi}, the FinGAN architecture consists of three modules: the generator, the discriminator, and the AuxNet. The generator receives a concatenated input of a Gaussian random vector and a 2-D RP vector to produce synthetic RSS samples. The discriminator takes either real or generated RSS samples and provides two outputs: one output with a sigmoid activation function indicating authenticity \cite{zhai2023sigmoid}, and the other one connects to the AuxNet. Assuming factored normal distributions for the $x$- and $y$-axis RPs, AuxNet predicts two means and two standard deviations; the latter are constrained to positive values using the Softplus activation function \cite{zamora2019adaptive}. The generator takes as input a 2-D RP vector and a 64-dimensional Gaussian noise vector, processed through 1D convolutional layers (256, 128, 64 filters) followed by a fully connected layer to produce 6-dimensional RSS outputs. Leaky rectified linear unit (ReLU) activations are applied after each convolution. The discriminator consists of 1D convolutional layers (128, 64, 16 filters) with standard ReLU activations. Its AuxNet branch includes a fully connected layer with 128 units, a ReLU activation, and a batch normalization layer, serving as a sub-network for auxiliary tasks. The FinGAN training alternates between updating the generator and discriminator, with one fixed while the other is optimized, as shown in the right side of Fig. \ref{fig:FinGAN_archi}. To ensure a fast convergence of the proposed algorithm, RSS data is standardized to zero mean and unit variance. Both networks are trained with Adam optimizers using a learning rate of $1 \times 10^{-4}$.

\section{Experimental Results}
\label{sec:Experimental Results}

\subsection{Dataset Details}
The real-world RSS samples were collected in a $30$ m $\times$ $10$ m  indoor environment with 6 RUs and 40 RPs, yielding 15,244 measurements \cite{11165772}. The dataset was split by RPs: 50\% (20 RPs, 7,666 samples) for training, 37.5\% (15 RPs, 5,582 samples) for validation, and 12.5\% (5 RPs, 1,996 samples) for testing. Since the validation and testing datasets are excluded from training, this split also enabled direct evaluation of FinGAN’s ability to generate RSS at unseen locations. Next, we define the performance metrics for analysis.

\subsection{RSS Generation Error}
\begin{figure}
    \centering
    \includegraphics[width=\columnwidth, height=0.35\columnwidth]{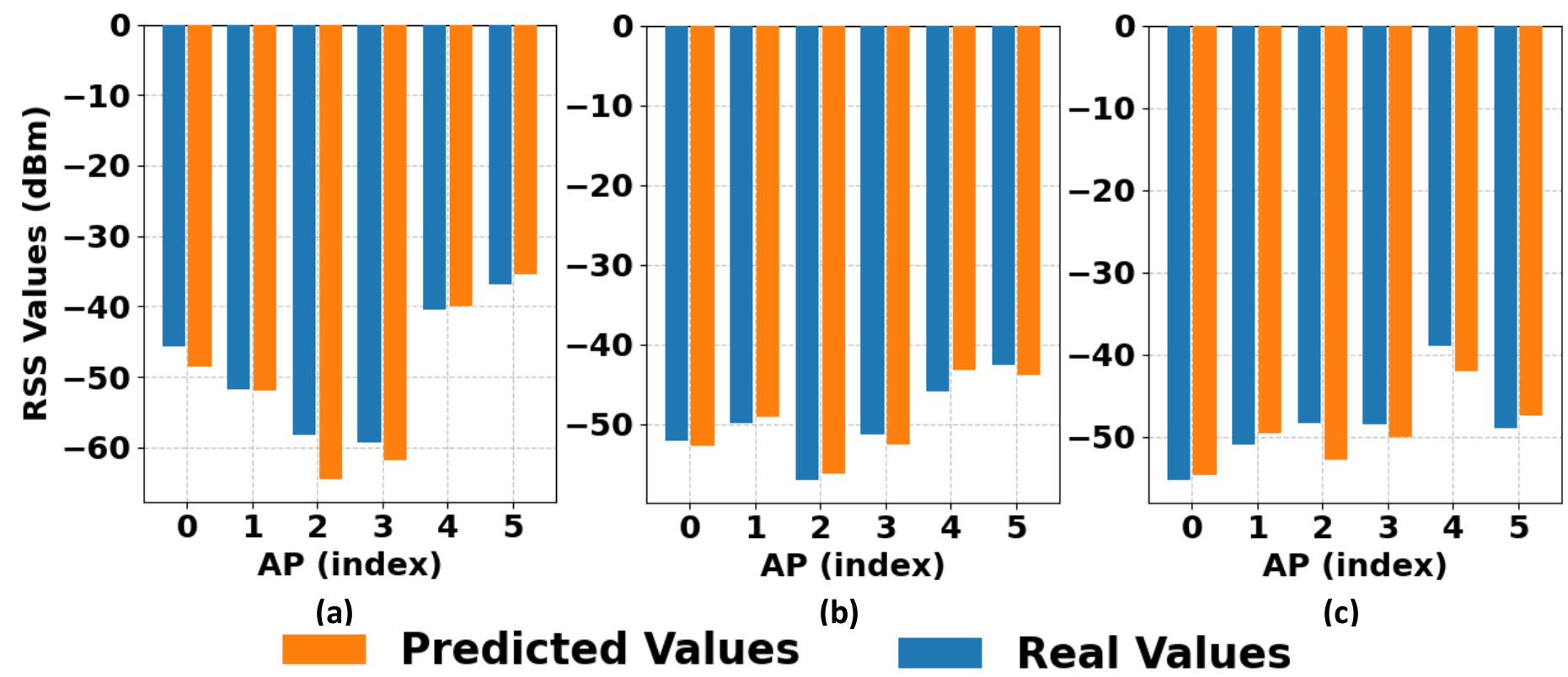}
    \caption{Comparisons of generated and ground-truth RSS samples from the proposed G-Te.}
    \label{fig:te_barcomp}
    \vspace{-10pt}
\end{figure}

To evaluate the performance of the proposed FinGAN, the average RSS error between the ground-truth and generated RSS samples corresponding to different RPs is calculated by using the root mean squared error (RMSE). As noted earlier, RSS measurements exhibit considerable variation even at fixed locations. To mitigate this, the RSS error is computed using the mean values of both ground-truth and generated samples at each RP, rather than averaging the RMSE across individual samples. For the 15 RPs in the validation set (G-V), the average element-wise error of the mean RSS values is 4.59 dBm. For the FinGAN testing set (G-Te), FinGAN-generated RSS samples at the RPs are compared against the corresponding ground-truth values. A representative subset is shown in Fig.~\ref{fig:te_barcomp}, with the associated RSS errors summarized in Table~\ref{tab:barcomp2}. Following the same methodology, the overall average element-wise error is 3.65 dBm. Both quantitative and qualitative analysis confirm that the proposed FinGAN effectively captures the underlying relationship between RPs and RSS distributions, enabling it to generate RSS samples that closely align with the true signal patterns associated with given locations.

\begin{table}[t]
\caption{RSS Error values of generated and ground-truth RSS samples at different RPs from the testing dataset.}
\setlength{\tabcolsep}{9.8pt}
\begin{tabular*}{\columnwidth}{|c|c|c|c|}
\hline
\textbf{index}          & (a)          & (b)         & (c)         \\ \hline
\textbf{RP (m)} & (4.61, 3.72) & (6.53, 4.37) & (6.03,4.35) \\ \hline
\textbf{RSS Error (dBm)}     & 3.01         & 1.38        & 2.43         \\ \hline
\end{tabular*}
\label{tab:barcomp2}
\vspace{-10pt}
\end{table}

\subsection{Localization Performance}
\label{subsec:loc}

To evaluate the reliability of localization with FinGAN-generated samples, we design a deep neural network (DNN) localizer composed mainly of fully connected layers and leaky ReLU activations (Fig.~\ref{fig:FinGAN_archi}, bottom left). The DNN is trained with the mean squared error (MSE) loss, while localization error is assessed using the RMSE.

Five training datasets are considered: (i) the FinGAN training set (G-T); (ii) G-T extended with G-V; (iii) G-T combined with Gen-V, the FinGAN-generated counterpart of G-V; (iv) G-V; and (v) Gen-V. The testing set in all cases is G-Te. As shown in Table~\ref{tab:locacc}, compared to G-T alone, G-T + G-V reduces localization error by 18.97\%, while G-T + Gen-V achieves a comparable 18.10\% reduction. Furthermore, when used independently, Gen-V performs on par with G-V, confirming that synthetic samples can effectively substitute for real measurements. To assess applicability of the proposed method, we test convolutional neural network (CNN) \cite{8662548} and $k$-nearest neighbour (KNN) \cite{10090664} localizers. For the CNN, the output is adapted from classification to $x$–$y$ coordinate regression, with MSE loss. KNN follows its standard setup. Across all three methods, hybrid or fully generated datasets achieve performance comparable to measured datasets, demonstrating that FinGAN-generated RSS preserves location-discriminative features and enables reliable localization.

\begin{table}[t]
\centering
\caption{Comparison of the localization error of Different Datasets by Different Localization Approaches.}
\setlength{\tabcolsep}{6.7pt}
\begin{tabular*}{\columnwidth}{|c|c|c|c|c|c|}
\hline
\multicolumn{1}{|c|}{\textbf{Method}} & \multicolumn{5}{c|}{\textbf{DNN}} \\ \hline
\textbf{Dataset} & G-T & G-T + G-V & G-T + \textbf{Gen-V} & G-V & \textbf{Gen-V} \\ \hline
\textbf{Values (m)} & 1.16 & 0.94    & 0.95       & 1.12 & 1.14      \\ \hline
\multicolumn{1}{|c|}{\textbf{Method}} & \multicolumn{5}{c|}{\textbf{CNN} \cite{8662548}} \\ \hline
\textbf{Dataset} & G-T & G-T + G-V & G-T + \textbf{Gen-V} & G-V & \textbf{Gen-V} \\ \hline
\textbf{Values (m)}    &  1.06   &   0.88   & 0.90       &   1.02    & 1.06      \\ \hline
\multicolumn{1}{|c|}{\textbf{Method}} & \multicolumn{5}{c|}{\textbf{KNN} \cite{10090664} } \\ \hline
\textbf{Dataset} & G-T & G-T + G-V & G-T + \textbf{Gen-V} & G-V & \textbf{Gen-V} \\ \hline
\textbf{Values (m)}&   1.23   &  1.16  & 1.17  &   1.14   &  1.15    \\ \hline
\end{tabular*}
\label{tab:locacc}
\vspace{-10pt}
\end{table}

\subsection{Ablation Study}

To assess the roles of the AuxNet module and the mutual information maximization mechanism, an ablation study is conducted. In this variant, both components are removed, leaving training to rely solely on adversarial loss. This simplified model is denoted lite-FinGAN, with all other architectural and training settings kept identical for fair comparison. Performance is evaluated in terms of average RSS prediction error and localization error. Under the G-T training scenario, lite-FinGAN yields RSS prediction errors of 9.05 dBm (Gen-V) and 8.28 dBm (Gen-Te), both substantially higher than those of the full model. For localization, combining measured RSS with lite-FinGAN data produces an error of 1.12 m using a DNN localizer, which is 0.17 m worse than FinGAN. These results highlight the critical role of AuxNet and mutual information maximization in improving RSS synthesis. Together, they enable FinGAN to capture richer structural dependencies, enhancing generalization beyond element-wise reconstruction and yielding more robust modelling of complex effects such as environment-induced signal bias.

\subsection{Performance Benchmarking}
To quantify the similarity between the real and generated RSS samples, inspired by the Fr\'echet Inception Distance (FID) \cite{NIPS2017_8a1d6947}, the Fréchet Distance (FD) metric is applied. Let $S$ denotes the total number of RPs, and let $(\mu_{tru}^{(s)},\Sigma_{tru}^{(s)})$ and $(\mu_{gen}^{(s)},\Sigma_{gen}^{(s)})$ denote the sample means and covariances of the ground ground-truth and generated RSS at RP $s$, respectively. The average FD is calculated as follows:
\begin{equation}
\begin{split}
    \mathrm{FD} = {}& \frac{1}{S}\sum\limits_{s=1}^{S}(||\mu_{tru}^{(s)} - \mu_{gen}^{(s)}||^{2} \\ {}& + Tr\bigl(\Sigma_{\mathrm{tru}}^{(s)} + \Sigma_{\mathrm{gen}}^{(s)}
       -2\sqrt{\Sigma_{\mathrm{tru}}^{(s)}\,\Sigma_{\mathrm{gen}}^{(s)}}\bigr),\\  
\end{split}
\end{equation}
where the trace term $Tr(\Sigma_{tru} + \Sigma_{gen} - 2\sqrt{\Sigma_{tru}\Sigma_{gen}}))$ succinctly measures differences in the distributional spread (covariance structures) of the RSS vectors. Since FD is sensitive to feature scaling and was originally designed for image embeddings in uniform feature spaces, both real and generated RSS samples are standardized using statistics from the training dataset. This ensures that FD reflects the distributional alignment between real and synthetic data, with lower values indicating closer agreement.

\begin{table}[t]
\centering
\caption{Comparison of the Quality of Synthetic Data from Different Approaches.}
\setlength{\tabcolsep}{4.1pt}
\begin{tabular*}{\columnwidth}{|c|c|c|c|c|c|}
\hline
\textbf{Method}   & Measured                 & FinGAN  & \begin{tabular}[c]{@{}c@{}}GPR-GAN \\ \cite{zou2020robot} \end{tabular} & \begin{tabular}[c]{@{}c@{}}cGAN-LSTM \\ \cite{10443392} \end{tabular}   &  \begin{tabular}[c]{@{}c@{}}cGAN \\ \cite{boulis2021data} \end{tabular}  \\ \hline
\multicolumn{1}{|c|}{\textbf{Dataset}} & \multicolumn{5}{c|}{\textbf{Gen-V}} \\ \hline
\textbf{\begin{tabular}[c]{@{}c@{}}RSS Error\\ (dBm)\end{tabular}}& N/A & \textbf{4.59} & 7.49   &   4.85 & 5.21\\ \hline
\textbf{FD} & N/A & \textbf{4.46} & 11.73  & 7.66 &  8.21\\ \hline
\multicolumn{1}{|c|}{\textbf{Dataset}} &\multicolumn{5}{|c|}{\textbf{Gen-Te}} \\ \hline
\textbf{\begin{tabular}[c]{@{}c@{}}RSS Error\\ (dBm)\end{tabular}} & N/A & \textbf{3.65} & 7.54   &    4.17 & 4.85\\ \hline
\textbf{FD} & N/A & \textbf{ 4.08} & 13.38   &  7.69  & 8.17 \\ \hline
\multicolumn{1}{|c|}{\textbf{Dataset}} &\multicolumn{5}{|c|}{\textbf{Gen-V + G-T}} \\ \hline
\textbf{\begin{tabular}[c]{@{}c@{}}Localization\\ Error (m)\end{tabular}} & 0.94 & \textbf{0.95} & 1.00   &  1.06 & 1.08\\ \hline
\end{tabular*}
\label{tab:compbar}
\vspace{-10pt}
\end{table}

To benchmark FinGAN, several GAN-based models previously applied to RSS augmentation are compared with it. GPR-GAN \cite{zou2020robot} models the distribution of RSS conditioned on RPs. For fairness, a cGAN-based method \cite{boulis2021data} and a cGAN enhanced with LSTM modules \cite{10443392}, both conditioned on RP attributes. Table~\ref{tab:compbar} reports results on Gen-V and Gen-Te. Localization errors are computed using hybrid datasets of G-T with Gen-V, following Section~\ref{subsec:loc}. FinGAN achieves the lowest average RSS error and FD across both sets: 4.59 dBm and 4.46 on Gen-V, and 3.65 dBm and 4.08 on Gen-Te. For localization, the hybrid set G-T+Gen-V yields the best accuracy of 0.95 m. These results show that FinGAN generates high-fidelity and diverse RSS data by maximizing RP information, leading to more accurate generation and superior localization performance over state-of-the-art GAN-based methods. In contrast, GPR-GAN inherits the modeling assumptions of GPR, which may limit its ability to capture complex RSS–RP relationships. Similarly, cGAN-based approaches rely on strict label consistency, which can constrain the diversity of generated RSS samples.

\begin{table}[t]
\centering
\caption{Comparison of the Quality of Synthetic Data from Different Approaches under Other Different Scenarios}
\setlength{\tabcolsep}{4.1pt}
\begin{tabular*}{\columnwidth}{|c|c|c|c|c|c|}
\hline
\textbf{Method}      & Measured              & FinGAN  & \begin{tabular}[c]{@{}c@{}}GPR-GAN \\ \cite{zou2020robot} \end{tabular} & \begin{tabular}[c]{@{}c@{}}cGAN-LSTM \\ \cite{10443392} \end{tabular}   &  \begin{tabular}[c]{@{}c@{}}cGAN \\ \cite{boulis2021data} \end{tabular}  \\ \hline
\multicolumn{6}{|c|}{\textbf{Dataset 1 \cite{feng2022analysis}}} \\ \hline
\multicolumn{1}{|c|}{\textbf{Dataset}} & \multicolumn{5}{c|}{\textbf{Gen-V$_{2}$}} \\ \hline
\textbf{\begin{tabular}[c]{@{}c@{}}RSS Error\\ (dBm)\end{tabular}} & N/A & \textbf{7.35} & 10.32 &   9.16   & 10.05\\ \hline
\textbf{FD} & N/A & \textbf{3.63} & 4.82  &  4.34 & 4.74 \\ \hline
\multicolumn{1}{|c|}{\textbf{Dataset}} &\multicolumn{5}{|c|}{\textbf{G-Te$_{2}$}} \\ \hline
\textbf{\begin{tabular}[c]{@{}c@{}}RSS Error\\ (dBm)\end{tabular}}& N/A & \textbf{7.33} & 9.74   &   7.92  & 8.31\\ \hline
\textbf{FD}& N/A & \textbf{3.48} & 4.04  & 3.60  &  3.81\\ \hline
\multicolumn{1}{|c|}{\textbf{Dataset}} &\multicolumn{5}{|c|}{\textbf{Gen-V$_{2}$ + G-T$_{2}$}} \\ \hline
\textbf{\begin{tabular}[c]{@{}c@{}}Localization\\ Error (m)\end{tabular}}& 1.98 & \textbf{2.05} &  2.36   &  2.29 &  2.35\\ \hline
\multicolumn{6}{|c|}{\textbf{Dataset 2 \cite{11016042}}} \\ \hline
\multicolumn{1}{|c|}{\textbf{Dataset}} & \multicolumn{5}{c|}{\textbf{Gen-V$_{2}$}} \\ \hline
\textbf{\begin{tabular}[c]{@{}c@{}}RSS Error\\ (dBm)\end{tabular}} & N/A & \textbf{5.49} & 13.63 &   9.61   & 12.21\\ \hline
\textbf{FD} & N/A & \textbf{7.40} & 35.56  &  18.16  &  21.74\\ \hline
\multicolumn{1}{|c|}{\textbf{Dataset}} & \multicolumn{5}{c|}{\textbf{Gen-Te$_{2}$}} \\ \hline
\textbf{\begin{tabular}[c]{@{}c@{}}RSS Error\\ (dBm)\end{tabular}} & N/A & \textbf{6.38} & 13.82  &  10.97    &  11.21\\ \hline
\textbf{FD} & N/A & \textbf{ 19.02} & 36.38  &  29.38  &  31.45\\ \hline
\multicolumn{1}{|c|}{\textbf{Dataset}} &\multicolumn{5}{|c|}{\textbf{Gen-V$_{2}$ + G-T$_{2}$}} \\ \hline
\textbf{\begin{tabular}[c]{@{}c@{}}Localization\\ Error (m)\end{tabular}}& 1.85 & \textbf{1.89} &  2.33   &  2.19 &  2.26\\ \hline
\multicolumn{6}{|c|}{\textbf{Dataset 3 \cite{yuen2022wi}}} \\ \hline
\multicolumn{1}{|c|}{\textbf{Dataset}} & \multicolumn{5}{c|}{\textbf{Gen-V$_{2}$}} \\ \hline
\textbf{\begin{tabular}[c]{@{}c@{}}RSS Error\\ (dBm)\end{tabular}} & N/A & \textbf{4.88} & 18.20  &  12.89 & 14.14\\ \hline
\textbf{FD} & N/A & \textbf{ 1.52} & 17.07  &   9.27  &  10.13  \\ \hline
\multicolumn{1}{|c|}{\textbf{Dataset}} & \multicolumn{5}{c|}{\textbf{Gen-Te$_{2}$}} \\ \hline
\textbf{\begin{tabular}[c]{@{}c@{}}RSS Error\\ (dBm)\end{tabular}} & N/A & \textbf{ 7.73} &  20.81 &   13.47  &  14.04 \\ \hline
\textbf{FD} & N/A & \textbf{ 3.07 } &  20.16  &  8.48  & 10.02  \\ \hline
\multicolumn{1}{|c|}{\textbf{Dataset}} &\multicolumn{5}{|c|}{\textbf{Gen-V$_{2}$ + G-T$_{2}$}} \\ \hline
\textbf{\begin{tabular}[c]{@{}c@{}}Localization\\ Error (m)\end{tabular}}& 1.36 & \textbf{1.38} &  2.51   &  2.62 & 2.74 \\ \hline
\end{tabular*}
\label{tab:Data2Comp}
\vspace{-10pt}
\end{table}

\section{Discussion}
\label{sec:Discussion}
\subsection{Analysis of Importance of Unseen RPs}
To further assess FinGAN’s capability in RSS database generation, a comparison between the generated data based on seen RPs and unseen RPs are conducted. Specifically, since G-V comprises 15 RPs and 5,582 samples, an equal number of synthetic RSS samples are generated using their corresponding RPs in G-T and trained model, denoted as Gen-T. Using a hybrid dataset of Gen-T and G-T reduced localization error to 1.00 m, compared to 1.16 m with G-T alone (a 13.79\% improvement). These results show that FinGAN can generate reliable RSS for both seen and unseen RPs, and that dataset diversity is more valuable than size alone in improving localization performance.

\subsection{Analysis on Information Maximization}
Recall that FinGAN is designed to maximize the mutual information between RPs and RSS, thereby enabling controllable RSS generation conditioned on RP values. In contrast, models such as the GPR approach in \cite{zou2020robot} focus solely on learning the spatial distribution of RSS. To examine the role of RSS-related latent codes, we replace the RP-based latent code with the average RSS values derived from G-T, and denote the generated samples as Gen-R. In this case, FinGAN loses controllability with respect to RPs, since the generation is no longer conditioned on RP information. Nevertheless, using a hybrid dataset of Gen-R and G-T yields a localization error of 1.03 m, corresponding to an 11.20\% improvement over G-T alone. This improvement arises from data augmentation, but the loss of controllability highlights that maximizing mutual information between RPs and RSS is critical for achieving both high-quality and RP-consistent RSS generation.

\subsection{Performance under Other Scenarios}
\label{sec:Data2}

To further evaluate FinGAN, we used three publicly available indoor RSS datasets \cite{feng2022analysis,11016042,yuen2022wi}. For each dataset, we re-partitioned the data into training (G-T$_2$), validation (Gen-V$_2$), and testing (Gen-Te$_2$) sets following the same strategy used for our collected dataset. Evaluation metrics follow the same protocol, using average RSS error, FD, and localization error from the DNN-based localizer. With G-T$_2$, localization errors are 2.26, 2.08, and 1.76 across the three datasets. Incorporating G-V$_2$ reduces these to 1.98, 1.85, and 1.36, while replacing G-V$_2$ with FinGAN-generated Gen-V$_2$ yields 2.05, 1.89, and 1.38. Comparisons with other state-of-the-art methods are summarized in Table~\ref{tab:Data2Comp}. FinGAN consistently achieves the lowest RSS error, FD, and localization error across all datasets, demonstrating both its adaptability and superiority.

\section{conclusion}
\label{sec:Conclusion}
This paper proposed FinGAN, a generative framework that synthesizes RSS samples by modelling the joint relationship between RPs and RSS values with mutual information maximization. FinGAN achieves localization accuracy comparable to real measurements and consistently outperforms existing methods in terms of RSS error, FD, and localization error across multiple datasets. These results demonstrate FinGAN as an efficient and adaptive solution for reducing the cost of manual RSS collection in practical localization systems. Future work will focus on integrating FinGAN with lightweight diffusion models to achieve a better trade-off between accuracy and computational efficiency.

\bibliographystyle{IEEEtran}
\bibliography{IEEEabrv,bibliography}

\end{document}